\documentstyle[pre,twocolumn,aps,epsf,floats]{revtex}

\begin{document}
\draft
\title{Modelling diffusion of innovations in a social network}  

\author{X. Guardiola$^{1}$, A. D\'{\i}az-Guilera$^{1}$, C. J. P\'{e}rez$^{1}$,
A. Arenas$^{2}$, and M. Llas$^{1}$}
\address{
	$^1$Departament de F\'{\i}sica Fonamental, Universitat de
 	Barcelona, Diagonal 647, E-08028
	Barcelona, Spain \\
	$^2$Departament d'Enginyeria Inform\`{a}tica, Universitat
Rovira i Virgili, Carretera Salou s/n, E-43006  Tarragona, Spain \\}
\maketitle

\begin{abstract}
A new simple model of diffusion of innovations in a social network with
upgrading costs is introduced. Agents are characterized by a
single real variable, their technological level. According to
local information agents decide whether to upgrade their level or not
balancing their possible benefit with the upgrading cost.
A critical point where technological avalanches display a
power-law behavior is also found. This critical point is characterized
by a macroscopic observable that turns out to optimize technological
growth in the stationary state. Analytical results supporting our
findings are found for the globally coupled case.  
\end{abstract}


There has recently been much interest in modeling social and
economical systems from a physical point of
view\cite{Mantegna:99a,Bouchaud:00a,Oliveira:99a}. Most of these
studies have fallen into two classes: statistical analysis of time
series and agent based microscopic models. Among the latter, most of
them have been proposed in order to mimic financial markets
behavior\cite{Eguiluz:00a,Lux:99a,Challet:00a}. Despite this, several
 authors have, on their turn, developed models to simulate other sort
of social behaviors such as the adoption of competing
products\cite{Goldenberg:00a}, innovation and
collaboration\cite{Arenas:00a,Arenas:01a} or group
decision-making\cite{Galam:00b}. The main goal of all these
models is to reproduce real world behavior while simplifying the theoretical
models retaining as less parameters as possible. 

Keeping this in mind, we have tackled the problem of diffusion of innovations
in a social network. In order to understand the complex behavior of
technology adoption dynamics one should consider how the stimulus for
change spreads by gradual local interaction through a social
network. Most of the times, these ``waves'' of change come in terms of
intermittent bursts separating relatively long periods of quiescence,
in other words, the system exhibits ``punctuated equilibrium''
behavior. Certainly some technologies, like cellular phones or VCR's, seem to
lurk in the background for years and then suddenly explode into mass 
use\cite{Krugman:96a}.

There are two main mechanisms involved in the diffusion of innovations
in a social network that any mathematical model should take into
account. On the one hand, there is a pressure for adopting
a new product or technology coming from marketing campaigns and mass media. These external processes
are essentially independent of the social network structure and one
can view their effects as a random independent process on the
individuals (hereafter called {\em agents}). On the other hand, there
is the influence of the surrounding agents who define the social network.
Once an agent decides to adopt a new technology, those who are
in contact with him can evaluate the new payoff the agent has got from
acquiring the new technology and compare it with their current
benefits. This propagating mechanism stands for interpersonal,
such as word of mouth, communication processes.  By balancing the
payoff increment with the associated upgrading cost, they may decide
to adopt, or not, the new technology. In this way, the local flux of
information plays a key role in diffusing new products. It is
important to notice that we are not considering any compatibility
constraint among the agents. Links only account for the flux of information
among agents who decide to take an action or another for their
exclusively own benefit.     

In this letter we propose a simple model of diffusion of technological
innovations with costs. In the simplest version of the model, a
population of $N$ agents lie in a one-dimensional chain with periodic
boundary conditions. Each agent $i$ is characterized by the real variable
$a_i$. This variable stands for their technological level, that is,
the higher $a_i$, the more advanced (technologically speaking) he is
. We will assume that the payoff that an agent receives from
possessing a certain technological level is simply proportional to
it. The model is then simulated as follows: 

(i) at each time step, a randomly
selected agent $a_i$ updates his technological level
\begin{equation}
a_i\rightarrow a_i+\Delta_i,
\end{equation}
where $\Delta_i$ is a random variable exponentially 
distributed with mean $\lambda$, that is, $p(\Delta) =
e^{-\Delta/\lambda}/\lambda$. This driving process accounts for
the external pressure that may lead to a spontaneous new technology
adoption by any of the population agents. In all
numerical simulations shown in this letter we have used
$\lambda=1/2$. However, all results are robust against other noise
choices, as long as they have a finite variance. 

(ii) all agents
$j\epsilon\Gamma(i)$ ($\Gamma(i)$ being the set of neighbors of agent
$i$) decide whether they also want to upgrade or not, according to 
the following rule 

\begin{equation}
a_i - a_j \geq C \Rightarrow a_j \rightarrow a_i,
\end{equation}
where $C$ (cost) is a constant parameter that stands for the price an
agent must pay in order to upgrade his technology as well as his personal
``resistance'' to change. 

(iii) if any $a_j$ has decided to also
upgrade his level, we let their neighbors also choose
whether to upgrade or not. This procedure is repeated until no one
else wants to upgrade, concluding a {\em technological avalanche}. 
Whenever an agent $a_i$ decides to
upgrade, their neighbors become aware of the new technology and
balance the profit they may obtain in case of also adopting it
($a_i-a_j$) with its cost $C$. It may well happen that if the technological
innovation spontaneously adopted by the seed of the avalanche is high
enough compared with the cost, the avalanche may end up spanning a large portion of the population. 

\begin{figure}
\label{perfil}
\centerline{
        \epsfxsize= 8.0cm
        \epsffile{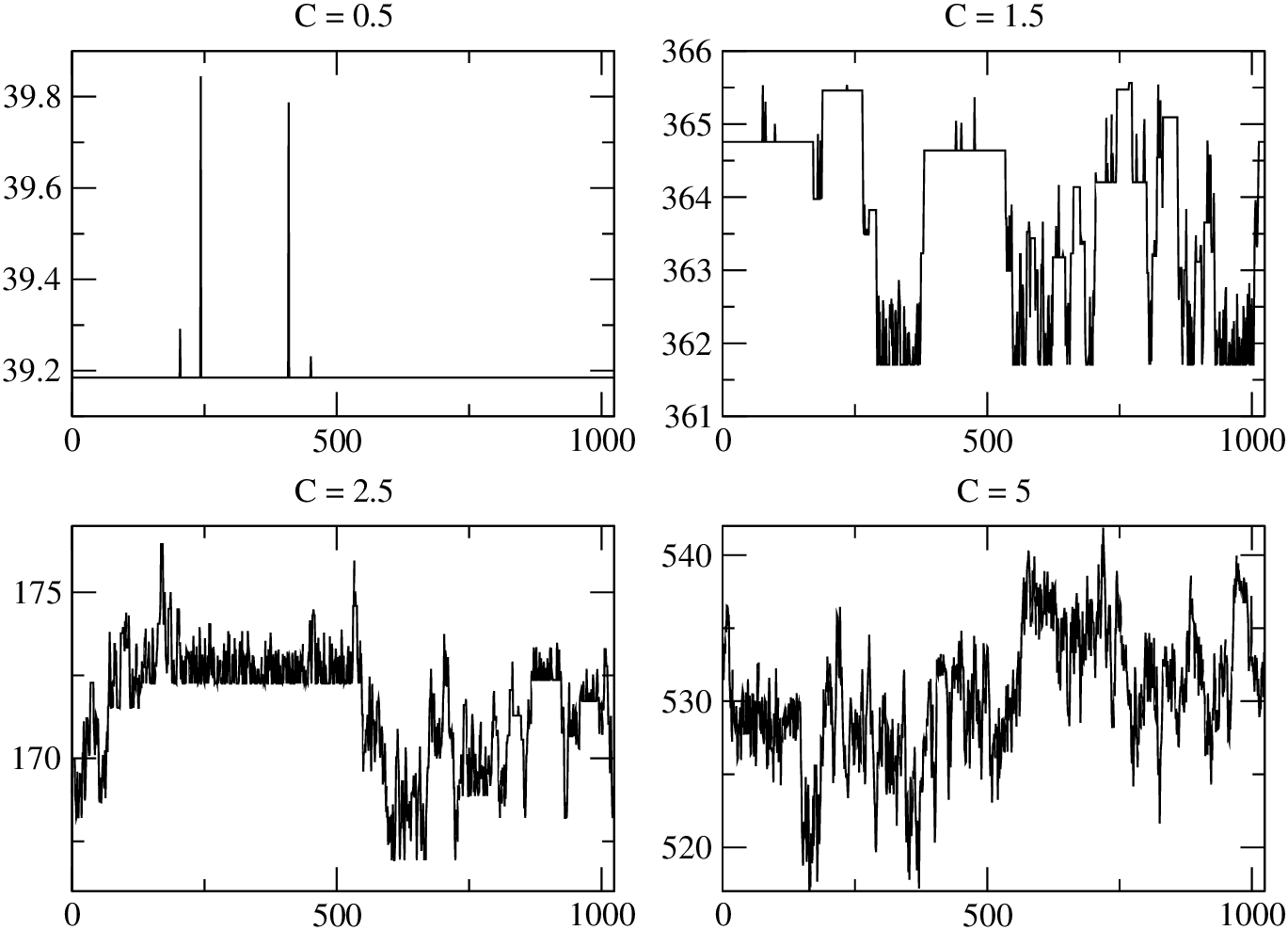}}
        \caption{Technological profiles ($N=1024$) for several values
        of $C$ in the stationary state. For $C=0.5$ the profile is
        almost flat (synchronized state) with everybody sharing the
        same technology. As we increase $C$ the profile gets noisier
        and plateaus (agents that, at some point have shared the same
        technology) become less common. For $C=5$ the technological
        profile is very random.}
\end{figure}

According to the cost value $C$ it is possible to distinguish several regimes.
In Fig. 1 we can see some examples of the technology profile (the
interface defined by the technology level of all agents) for several
values of the cost $C$.
For $C\ll1$, once there is an external random update, a system size
avalanche is immediately triggered so that all agents end up sharing
the same technological level, or in other words, the system is always
in an almost synchronized state. For values of $C\gg1$ upgrading is so
expensive that agents do not care about their neighbors technology,
and large avalanches are not triggered any more (almost all avalanches are
of size $1$). In this regime the technological profile is quite rough
(actually, in the limit $C\rightarrow\infty$ we should recover the
random deposition model\cite{Barabasi:95a}). 
In between these two regimes, there is a region showing a
rich dynamics where one finds technological avalanches of all
possible sizes. Actually, for some values of $C$ the probability
density of having an avalanche of size $s$ shows a power-law behavior 

\begin{equation}
P(s)\sim s^{-\tau}.
\end{equation}
Fig. 2 shows $P(s)$ for several system sizes and $C=3$.
The appearance of power-law distributed quantities is usually related
to the existence of some critical point. Nevertheless,
it is difficult to locate the critical point by looking at $P(s)$
since finite-size effects provide a whole region of the parameter
space where $P(s)$ behavior is compatible with a power-law. Actually,
the same problem appears in some SOC\cite{Jensen:98a} models and the
question of whether there is critical point or a whole critical region
in some parameter space has been largely debated\cite{Lise:01a}. We need
another signature of criticality that may help us in locating the
critical point.

\begin{figure}
\label{powerlaw}
\centerline{
        \epsfxsize= 8.0cm
        \epsffile{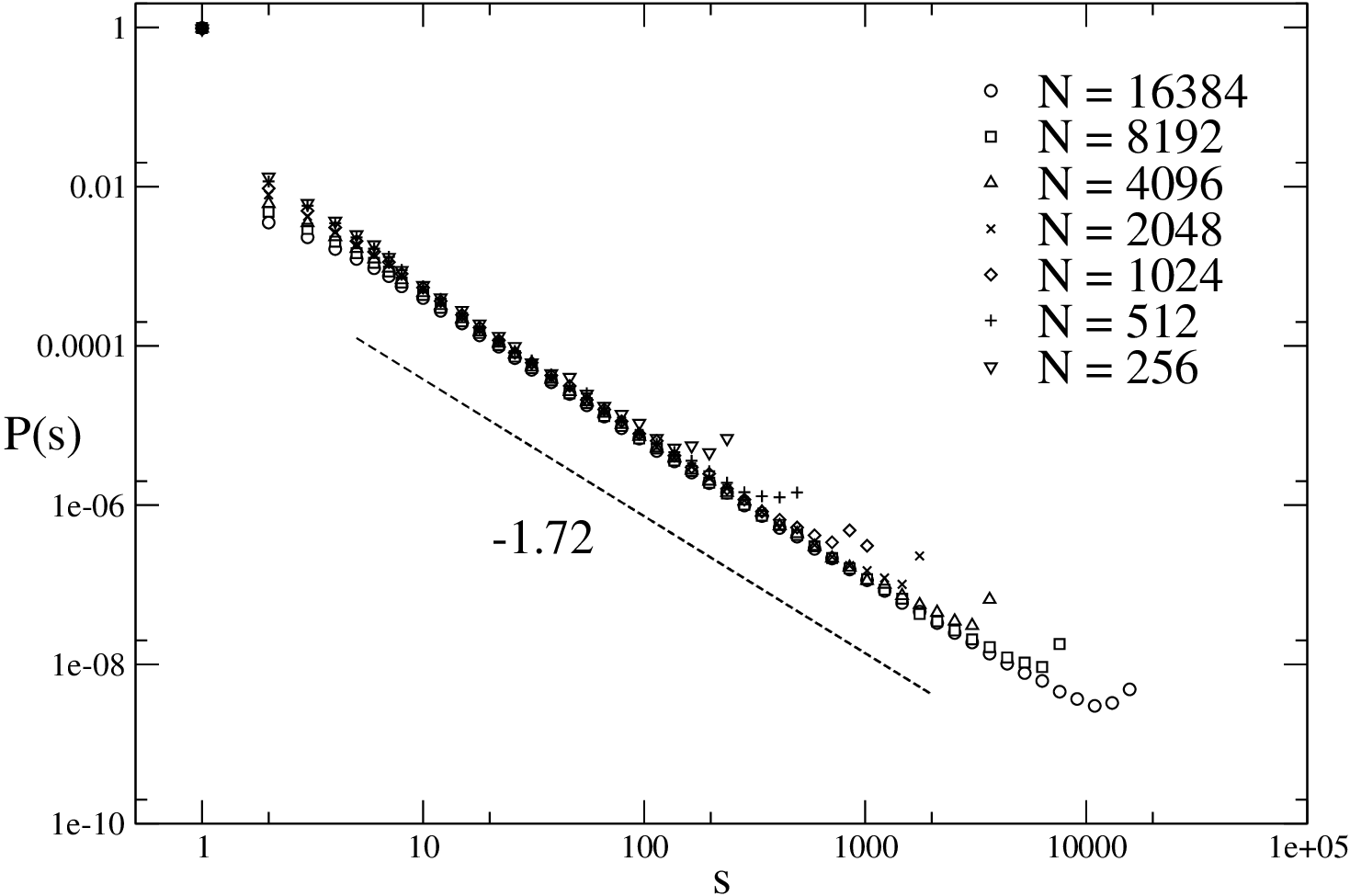}}
        \caption{Plot of the probability density of having a
        technological avalanche of size $s$ for $C=3$ in log-log
        scale. We find that
        $P(s)\sim s^{-\tau}$ with $\tau=1.72$. The peaks at the
        end of each curve are due to finite size effects. Numerical 
	simulations have been averaged over $10^8$ avalanches}
\end{figure}

A possible answer comes from the social interpretation of the
model. Social science researchers usually work with aggregated data such as
the adopting curve\cite{Valente:96a}, that is, the evolution in time
of the total number of people who adopts a certain product or
technology. Analogously, in our model we can set an arbitrary
threshold $a_{th}$ and then calculate how many agents posses a
technology $a > a_{th}$ as the upgrading process goes on. Let us label
$\phi$ the fraction of agents with $a > a_{th}$. Fig. 3
shows three adopting curves for three different values of
$C$. For $C = 1.25$ large avalanches (made of a lot of agents acquiring
the same new product or technology) are
triggered. In this way, the technological profile
advances uniformly and, the system must pay a lot of
costs. This situation is clearly inefficient. In the plot this is
reflected by the fact that the curve for $C=1.25$ is the last one to
reach $a_{th}$ (at least one of the agents), but once it begins the
whole population crosses $a_{th}$ very fast (because of the uniform 
advance).   
On the other hand, for $C=3.5$ very few avalanches 
are triggered, meaning that the profile grows in a
very non-uniform way, and its fluctuations are quite important. That
is why the $C=3.5$ curve begins crossing the threshold $a_{th}$ 
earlier than the
case $C=1.25$. However, it takes much more time the
whole population to cross the threshold and it is clearly inefficient in
terms of how many times a cost is paid.
As result, there is an intermediate value of $C$ ($C=2.5$ in the plot)
where this weighted growth process is optimized and the $\phi$ curve is
always greater than for larger and smaller cost values. That corresponds to
the critical region, where avalanches of all possible sizes are
triggered. In other words, there are some intermediate $C$ values
that let the population reach a given average technological level with
a minimum number of upgrades (and their associated costs). Therefore, we
can speak of an efficient cost region leading to an optimal growth rate.  

We can quantify this effect by computing the so-called
{\em mean velocity of progress}\cite{Arenas:00a} defined as 
ratio of the total technology advance and the total number of
upgrades. It can also be computed as 
$\rho = \left<H\right>/\left<s\right>$, where $\left<H\right>$
stands for the average total technological advance induced by an avalanche
(the interface area increment caused by an avalanche) and
$\left<s\right>$ is the average avalanche size. This quantity,
$\rho$, gives an idea of how fast the technological profile grows.
Fig. 4 shows several plots of $\rho$ against $C$ for several
system sizes. The first thing one can see is that $\rho$ has a maximum
for an intermediate value of $C$. Moreover,
$\rho_{max}$ scales with the system size as $\rho_{max}\sim N^{0.20(1)}$,
diverging in the thermodynamic limit $N\rightarrow\infty$. The
location of $\rho_{max}$ allows us to define the {\em finite size}
critical point of the model $C_cN$. A proper and detailed
characterization of this critical point will be published elsewhere.
It is also possible to exactly calculate the assymptotic value of
$\rho$. For $C\ll 1$, almost all avalanches are of the size of the
system $\left<s\right>\sim N$, and the total advance induced by them
is, in average, $N\lambda$. Therefore,
$\rho\rightarrow\lambda$. Moreover, for $C\gg 1$ the avalanches are of
unit size $\left<s\right>\sim 1$ and advance $\lambda$, so that
$\rho\rightarrow\lambda$ as well.
\begin{figure}
\label{phi}
\centerline{
        \epsfxsize= 8.0cm
        \epsffile{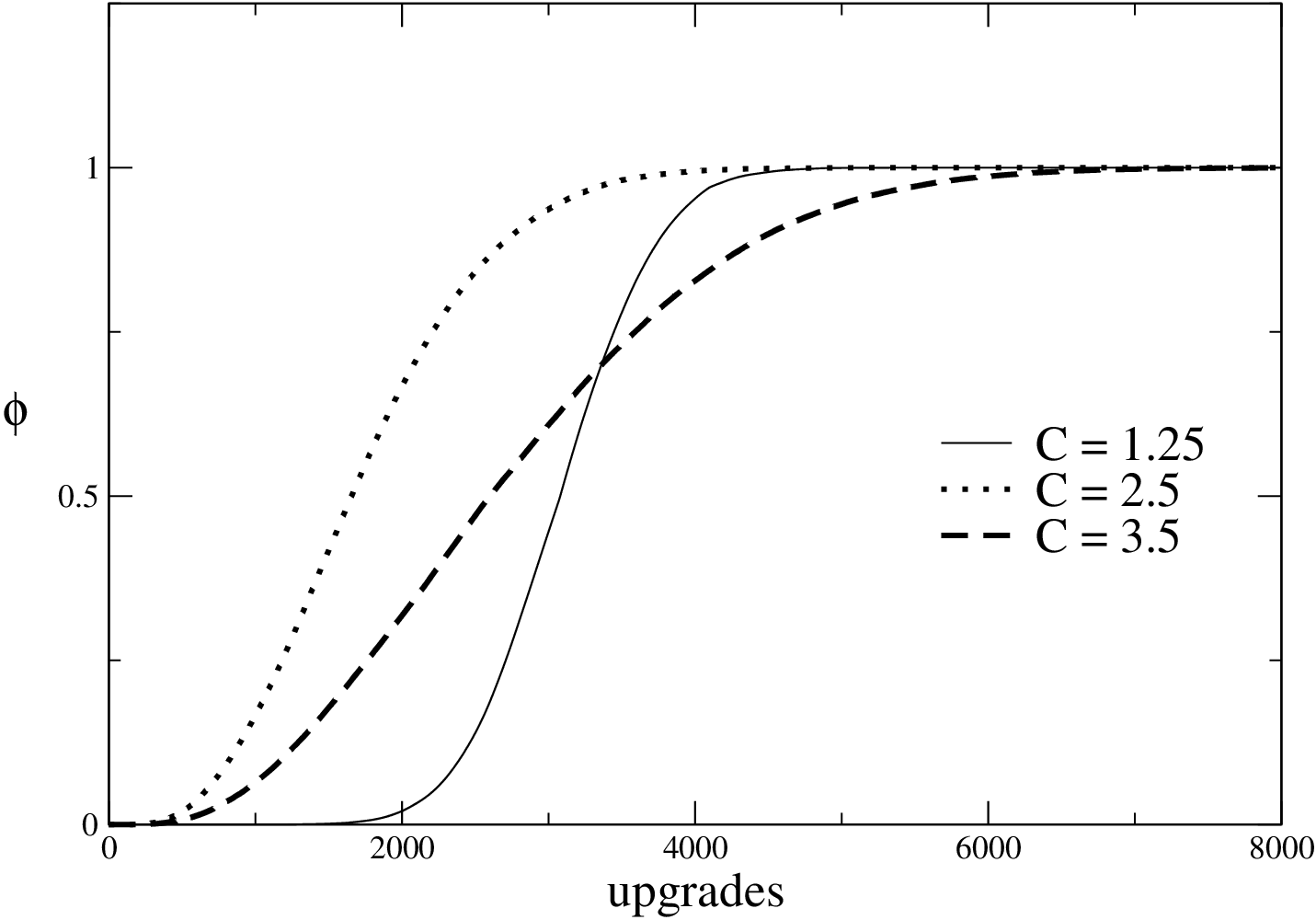}}
        \caption{Evolution of the fraction of agents with $a >
        a_{th}=3$ for $N = 1024$ and several values of $C$ as the
        number of upgrades increases. Numerical simulations have been
	 averaged over $1000$ initial configurations. In this example the
        curve for $C = 2.5$ is always above the others.}
\end{figure}

In the view of all this, one can assert that it is near the critical
point where the technological profile grows more efficiently. This
leads to the following paradoxal result: upgrading costs should be
neither cheap nor expensive in order to have an optimal technological
growth. Obvioulsy, our concept of efficiency is related to the number
of times a cost is paid, that is, from the point of view of the
population but not the companies who sell the products. Sellers will
always look for a scenario where agents acquire as many new products 
as often as possible.

In order to complete our study, we have also analysed the globally coupled
case, where some analytical results have been found. 

\begin{figure}[t]
\label{rho}
\centerline{
        \epsfxsize= 8.0cm
        \epsffile{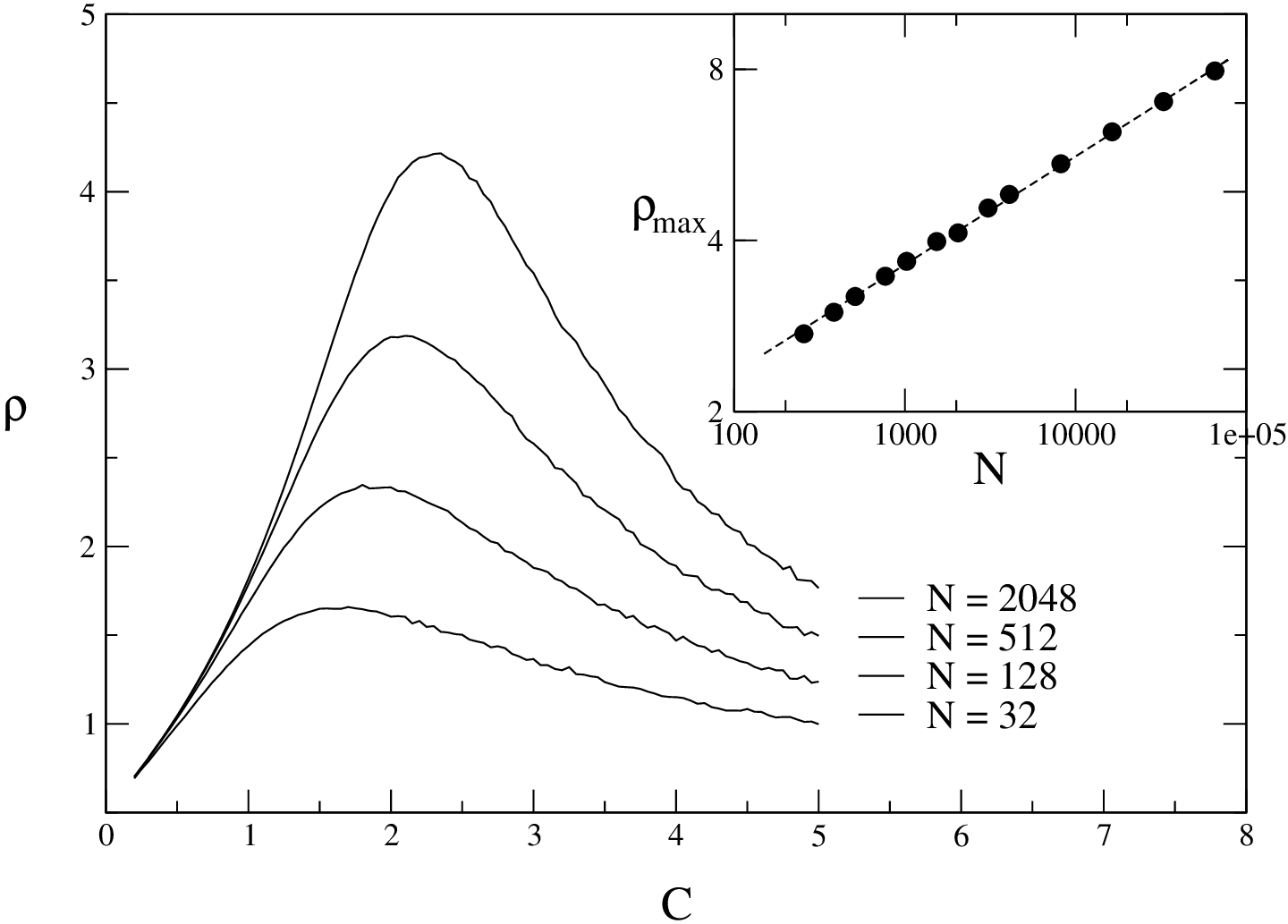}}
        \caption{$\rho$ as a function of $C$ for several system
        sizes. At the extremes
        $C\rightarrow 0$ and $C\rightarrow\infty$, the value of $\rho$
        goes to $<\Delta_i>=\lambda=1/2$. There is a peak,
        $\rho_{max}$, that diverges in thermodynamic limit
        $N\rightarrow\infty$. In the inset we plot $\rho_{max}$
        against system size $N$. Dashed line shows the fit $\rho_{max}\sim N^{0.20(1)}$.} 
\end{figure}

In the globally coupled version of the model information about the
technological level $a_i$ of {\em all} agents is available to {\em
any} agent. Now, agents technological level is confined in a band of
width $C$ since whenever there is a difference $a-a' \ge C$ between
any two agents, the one with the lowest level immediately adopts the
highest technological level. Moreover, the system still displays a
peak for the mean velocity of progress $\rho$ as Fig. 5
shows. What is, indeed, also quite amazing, is that the globally
coupled case also has also a power-law avalanche probability
distribution $P(s)$ at the efficient region (Fig. 5).

In order to give an estimation of $\rho$
in the stationary state, let us make some mean-field assumption (that
is, restricting to average values and neglecting fluctuations). Let us
 assume that the agents technological levels are uniformly
distributed over the band of width $C$, so that there is a density of
levels $N/C$. In order to keep things simple we also assume that
random spontaneous updates are of fixed size $\lambda$.  
Then, there is going to be an avalanche whenever any of the agents having a
technological level $a\in[C-\lambda,C]$(where the origin has been set
at the base of the band) decides to, spontaneously, adopt a new
technology. This will happen with a probability $\lambda/C$ every time
step. The agents involved in such avalanche will be those
who lie in the lowest region of the band $a\in[0,\lambda]$, and, in
average, half of them will take part in the avalanche, so that the
number of agents involved is $N\lambda/2C$. These agents will advance
their technological levels by $C$. Therefore, after $T$ time steps, in
average there will be $T\lambda/C$ avalanches and $T(1-\lambda/C)$
simple spontaneous updates. Now, we can calculate $H$, the global
technological advance after $T$ time steps, as well as $S$ the total
number of upgrades (spontaneous and induced by the avalanches)

\begin{equation}
H = \frac{T\lambda}{C}\left(\frac{N\lambda}{2C}C\right)
+T\left(1-\frac{\lambda}{C}\right)\lambda
\end{equation}
\begin{equation}
S = T + \frac{T\lambda}{C}\frac{N\lambda}{2C}.
\end{equation}
Then $\rho$ is imply given by the ratio $H/S$. In terms of the
adimensional variables $\rho/\lambda$ and $\mu\equiv\lambda/C$ we find
the relation

\begin{equation}
\rho/\lambda = \frac{\mu N/2 + 1-\mu}{1+\mu^2N/2}.
\end{equation}
This formula holds whenever $\lambda > C$, otherwise the above
assumptions are not valid, and one trivially finds that
$\rho=\lambda$. Notice that in the limit $C\rightarrow\infty$ we
also recover $\rho\rightarrow\lambda$. Fig. 5 shows a comparison 
between Eq. (6) and simulation data. Although the formula gives a
correct estimation for $C\rightarrow 0$ and $C\rightarrow\infty$,
there is some discrepancy near the peak of the $\rho$. A plausible
explanation for this is the existence of large fluctuations (as
Fig. (5) shows) so
that a mean-field approach only provides a crude estimation of $\rho$.

Now, it is possible to study the
asymptotic behavior of $\rho_{max}$ that results from maximizing
Eq. (6). We find that

\begin{equation}
\mu_{cN} = \frac{-2N-\sqrt{2N(N^2+2N-4)}}{N(2-N)}.
\end{equation}
In the thermodynamic limit $N\rightarrow \infty$,
$\mu_{cN}\rightarrow 0$ and $C_{cN}\rightarrow \infty$. Therefore, for the
globally coupled case the critical point goes to infinity and
$\rho_{max}$ diverges as $\rho_{max}\sim N$.

 In conclusion, we have presented a simple model of
diffusion of innovations in a social network displaying
rich dynamics ranging from global synchronization to critical
behavior. Costs are responsible of blocking the flux of information
over the network, but, at the same time, they are necessary to
guarantee an optimal growth of the technology profile. In order to
show this, we have computed the value of $\rho$, mean velocity of
progress, a quantity that is maximized at the critical point of the
model. We have also analytically solved a mean-field version of the
globally coupled case and showed the existence of a maximum value of $\rho$
that diverges in the thermodynamic limit. Also in this case, a power-law
avalanche distribution leading to a critical behavior has also been
found at the efficient region of the model. Therefore, one of the most
interesting things of our model is that all its most intriguing
features are qualitatively the same regardless of the systems connectivity.

We kindly acknowledge F. Vega-Redondo, J.
J. Ramasco, M. A. Rodr\'{\i}guez and J. M. L\'{o}pez 
for helpful comments. Financial support has been provided by DGES of
the Spanish Government through grant BFM2000-0626 and EU TMR Grant No
ERBFMRXCT980183. X. G. and M. Ll. also acknowledge financial support from the
Generalitat de Catalunya and the MCT of the Spanish government, respectively.

\begin{figure}[t]
\centerline{
        \epsfxsize= 8.0cm
        \epsffile{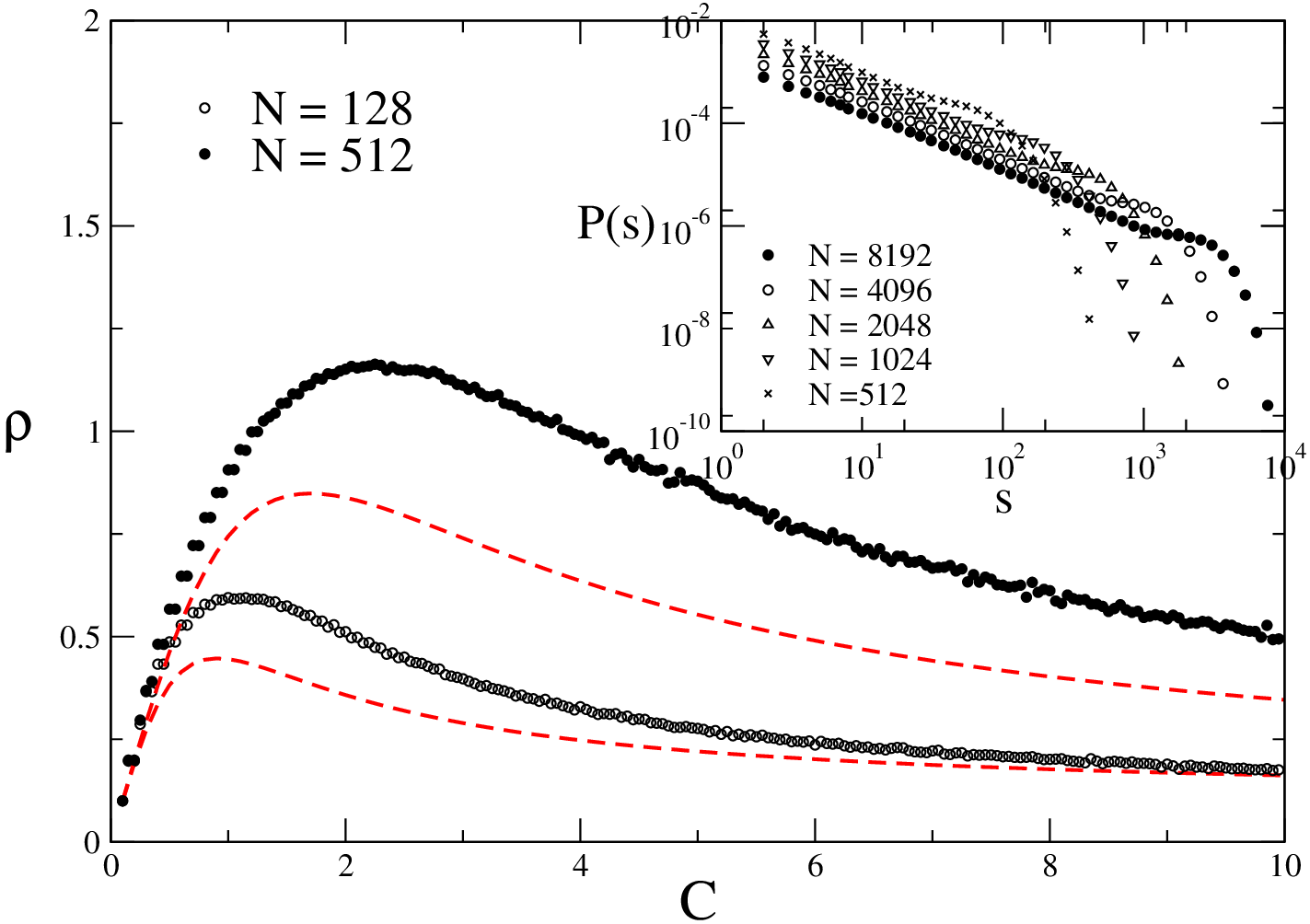}}
        \caption{$\rho$ for different system sizes as a function of
        $C$ for the globally coupled case. Dashed line stands
        for Eq. (6). In the inset we plot the avalanche probability
        distribution $P(s)$ around the peak of the $\rho$. For the
        simulations we have used $\lambda=0.1$.}
\end{figure}


\end{document}